\documentclass[twocolumn,aps,prl,reprint,showpacs,amsmath,amssymb]{revtex4-1}
\usepackage{epsfig}
\usepackage{graphicx}
\usepackage{dcolumn}
\usepackage{bm}
\usepackage{bookmark}
\usepackage{color}
\usepackage{ulem}

\begin{document}
\title{Kibble-Zurek mechanism in a one-dimensional incarnation of deconfined quantum critical point}
\author{Rui-Zhen Huang$^1$}
\author{Shuai Yin$^2$} \email{yinsh6@mail.sysu.edu.com}
\affiliation{$^1$Kavli Institute for Theoretical Sciences, University of Chinese Academy of Sciences, Beijing 100190, China}
\affiliation{$^2$School of Physics, Sun Yat-Sen University, Guangzhou 510275, China}

\date{\today}

\begin{abstract}
The conventional Kibble-Zurek mechanism (KZM) describes the driven critical dynamics in the Landau-Ginzburg-Wilson (LGW) spontaneous symmetry-breaking phase transitions. However, whether the KZM is still applicable in the deconfined quantum criticality, which is beyond the LGW paradigm, has not been explored. In this paper, we study the driven critical dynamics near a one-dimensional incarnation of deconfined quantum critical point between a ferromagnetic (FM) phase and a valance-bond-solid (VBS) phase. By investigating the dependence of the density of the topological defects on the driving rate, we verify the KZM in this Landau-forbidden critical point. Moreover, we find that both the FM and the VBS order parameters satisfy the finite-time scaling in the whole driven process. The effects of the emergent symmetry in the nonequilibrium dynamics are also studied.
\end{abstract}

\maketitle

{\it Introduction}.--- Fathoming universal properties of phase transitions is among the central issues in modern statistical mechanics and condensed matter physics~\cite{Wen,Fradkin,Sachdevbook}. A prevalent understanding of phase transitions is based on the Landau-Ginzburg-Wilson (LGW) paradigm~\cite{Landau,Wilson1974}, which shows that the continuous phase transition is characterized by the spontaneous symmetry breaking from a disordered phase to an ordered phase. However, recently this paradigm has been challenged by a series of examples, in which continuous phase transitions were found between two ordered phases, such as the phase transition between the Neel antiferromagnetic (AFM) phase and the valance-bond-solid (VBS) phase in (2+1)-dimension (D) Heisenberg spin magnets~\cite{Senthil1,Senthil2}. To understand these Landau-forbidden critical phenomena, a concept of deconfined quantum critical point (DQCP) was proposed~\cite{Senthil1,Senthil2}. The DQCP theory asserts that deconfined fractionalized particles emerge near the critical point and their fluctuations dominate the phase transition properties~\cite{Senthil1,Senthil2}. Although great efforts have been spent on examining the DQCP theory in various systems, ambiguities among the nature of the transition are still not completely clarified~\cite{Senthil1,Senthil2,Senthil3,SandvikDQCP,Lee19,Shao,Nahum1,Nahum2,Nahum3,Meng1,Zhang}.

In nature equilibrium state is the exception rather than the rule. In particular, near a critical point the equilibration time diverges owing to the critical slowing down. This has stimulated intensive investigations on the nonequilibrium critical dynamics in both classical and quantum systems~\cite{revqkz1,revqkz2}. Among them, the driven critical dynamics under external driving stands out remarkably, partly spurred by its potential application in quantum simulation and quantum computer~\cite{revqkz1,revqkz2}. For the driven dynamics in the LGW symmetry-breaking phase transitions, the celebrated Kibble-Zurek mechanism (KZM) provides a description of the generation of the topological defects and the scaling of their number after the quench~\cite{Kibble1,Zurek1}. While the KZM was originally proposed in cosmological physics~\cite{Kibble1}, it has been generalized to condensed matter phase transitions in classical and quantum systems~\cite{Zurek1,qkz1,qkz2,qkz3,qkz4,qkz5,qkz6,Matuszewski,del,Millis,Mukherjee}. Moreover, the KZM has been verified in various experiments~\cite{Monaco,Ulm,Pyka,Navon,Autti,Ko,Keesling}. Besides, recent theoretical and experimental studies also pay close attentions to the full scaling behavior in the whole driven process~\cite{Keesling,qkz7,qkz8,qkz9,Chandran,Yin}. For example, a finite-time scaling (FTS) theory gives a full scaling theory and shows that the critical dynamics near the critical point is dominated by the time scale induced by the external driving~\cite{Zhong1,Zhong2}. The FTS theory has been verified in the driven critical dynamics of the Rydberg atomic systems~\cite{Keesling}. Furthermore, these full scaling forms have been employed to numerically detect the critical properties in both classical and quantum phase transitions~\cite{Zhong1,Zhong2,Huang,Hu,HuangRZ,Sandvik05,Zhong3}.

However, for the DQCP, the nonequilibrium critical dynamics therein has not been explored. It is natural to ask whether the KZM and the FTS are still applicable in the DQCP. But directly studying on the real-time dynamics in $(2+1)$D DQCP is hindered by the lack of reliable method. For example, the quantum monte carlo fails as a result of the sign-problem~\cite{Sandvik,Liu}, while the tensor-network method still needs tremendous improvements to simulate the long-time dynamics~\cite{Verstraete,Jordan,Czarnik1,Czarnik2}.

In this paper, we take a detour and investigate the driven dynamics in a $(1+1)$D incarnation of DQCP~\cite{Jiang}, in which a Landau-forbidden continuous phase transition happens between a ferromegnetic (FM) phase and a VBS phase. By driving the system across its critical point, we find that the density of residual topological defects after the quench satisfies the KZM. In addition, we explore the evolution of the FM and the VBS order parameters and show that their scaling behaviors in the whole driven process can be described by the FTS theory. Moreover, since at the DQCP there are some emergent symmetries, which manifest themselves in the corresponding correlation functions~\cite{HuangRZDQCP}, we also study the FTS of the correlation functions. Although there are some differences between this $(1+1)$D DQCP and the $(2+1)$D DQCP~\cite{Senthil1,Senthil2,Jiang}, for example, the deconfined spinons already exist in the excited states of ordered phases on both sides of the critical point, because of the lack of confining potential, we take a first step in this issue and the remarkable similarities between them indicate that our conclusions can be generalized to the $(2+1)$D case with some proper modifications.

{\it Model and its static properties}.---  We begin our study with a quantum spin model proposed by Jiang and Motrunich~\cite{Jiang}. The Hamiltonian reads
\begin{eqnarray}
\begin{aligned}
  H=&\sum_i(-J_x S_i^xS_{i+1}^x-J_z S_i^zS_{i+1}^z) \\
      &+\sum_i(K_x S_i^xS_{i+2}^x+K_z S_i^zS_{i+2}^z), \label{modelham}
\end{aligned}
\end{eqnarray}
in which $S_i^{x(z)}$ is the spin-$1/2$ operator in $x(z)$ direction at $i$ site. Besides the translational symmetry, two $Z_2$ symmetries, i.e., $Z_2^x$ and $Z_2^z$, corresponding to the spin inversion in $x$ and $z$ directions, respectively, are respected in model~(\ref{modelham}).

With fixed parameters $K_x=K_z=1/2$ and $J_x=1$, a gapped FM phase, whose order parameter is $m\equiv \langle S_i^z\rangle$, appears for large $J_z$, while a VBS phase, whose order parameter is $\psi\equiv \langle {\bf S}_i\cdot{\bf S}_{i+1}-{\bf S}_{i+1}\cdot{\bf S}_{i+2}\rangle$, appears when $J_z$ is small. In particular, when $J_z=1$, the system is at the Majumdar-Ghosh point with an exact VBS ground state wave function. By tuning $J_z$, there is a direct phase transition occurring at $J_{zc}=1.4645$ between the FM and the VBS phases~\cite{HuangRZDQCP}. According to the LGW paradigm, this phase transition should be first-order, since the FM phase and the VBS phase break different symmetries. However, both theoretical and numerical results show that this is a continuous phase transition~\cite{Jiang,HuangRZDQCP,Roberts,Mudry,Sun}, in analogy to the $(2+1)$D DQCP~\cite{Senthil1,Senthil2}, in which the order parameters in two sides of the transition point can be regarded as different composites of fractionalized quasiparticles. Moreover, it was shown that the equilibrium universal behavior of this $(1+1)$D DQCP is described by the Luttinger liquid theory with continuously tunable Luttinger parameters~\cite{Jiang,HuangRZDQCP}. For the set of the parameters chosen above, the critical exponent $\nu$, defined as $\xi\propto |g|^{-\nu}$ with $g\equiv J_z-J_{zc}$ and $\xi$ being the correlation length, is $\nu\simeq 1.61$, and the dynamic exponent $z=1$. According to the self-duality of the parton description of this DQCP, the order parameter exponent $\beta$ in the ferromagnetic phase is identical with that in the VBS phase and $\beta\simeq 0.53$~\cite{HuangRZDQCP}.

Similar to the case in the $(2+1)$D DQCP~\cite{Nahum1,Nahum2,Nahum3,Meng1}, an emergent $O_\phi(2)\times O_\theta(2)$ symmetry arises at $J_{zc}$~\cite{HuangRZDQCP}. $O_\phi(2)$ corresponds to the rotation symmetry in the plane spanned by the FM operator and the VBS operator, and $O_\theta(2)$ corresponds to the rotation in the plane spanned by the FM operator in $x$-direction and the AFM operator in $y$-direction~\cite{HuangRZDQCP}. This emergent symmetry imposes strong constraint on the scaling of the correlation function~\cite{HuangRZDQCP}. For example, it was shown that the correlation of the $xy$-dimer, $G_\Gamma(r)\equiv (-1)^r\langle\Gamma_i \Gamma_{i+r}\rangle$ with $\Gamma_i$ being the $xy$-dimer operator defined as $\Gamma_i\equiv S_i^xS_{i+1}^y$, must satisfy $G_\Gamma(r)\sim 1/r^2$~\cite{HuangRZDQCP}.
\begin{figure}
	\includegraphics[angle=0,scale=0.25]{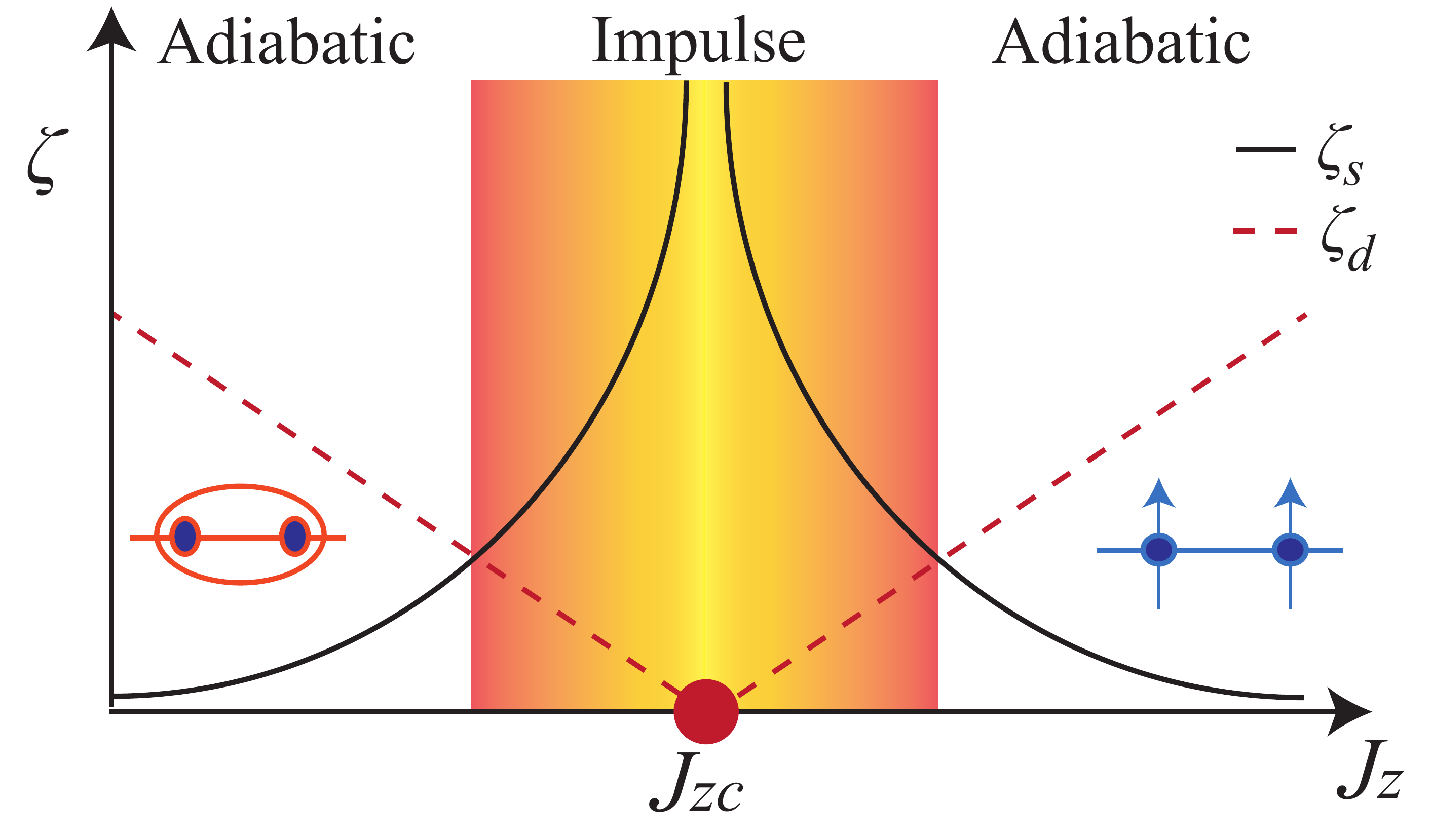}
	\caption{Schematic diagram of the KZM in $(1+1)$D DQCP. By tuning $J_z$, a direct continuous phase transition happens at $J_{zc}$ between the VBS phase (left) and the FM phase (right). This phase transition is beyond the LGW paradigm. We generalize the KZM into this phase transition. By comparing the relaxation time $\zeta_s$ and the inverse transition rate $\zeta_d$, the KZM separates the driven process into three regions: one impulse region and two adiabatic regions. In the adiabatic regions, $\zeta_s<\zeta_d$; while in the impulse region, $\zeta_s>\zeta_d$.}
	\label{fig1}
\end{figure}

{\it KZM in $(1+1)$D DQCP}.--- The KZM is a mechanism describing the production of the topological defects and their scaling behaviors after a quench across a critical point. We consider the driven dynamics by changing $g$ as $g(t)=Rt$. This linear quench imposes a driving rate $\zeta_d^{-1}\equiv |g^{-1}\frac{dg}{dt}|=t^{-1}$ on the system~\cite{revqkz1,revqkz2,Kibble1,Zurek1}. By comparing $\zeta_d$ and the intrinsic relaxation time scale of the system $\zeta_s$, which scales with the energy gap $\Delta$ as $\zeta_s\sim \Delta^{-1}$~\cite{Sachdevbook}, the KZM separates the whole driven process into three regions: one impulse region sandwiched by two adiabatic regions. In the initial adiabatic region, the system is far from its critical point and $\zeta_s<\zeta_d$. Thus, the system evolves adiabatically along the ground state. As the system gets closer to the critical point, the system enters the impulse region at $g=\hat{g}_1$ with $\zeta_s(\hat{g}_1)=\zeta_d$. In this region, $\zeta_s>\zeta_d$ and the adiabaticity breaks down. The KZM assumes that the system does not evolve in this region and the state remains the same as that at $g=\hat{g}_1$. Then, continuing driving pushes the system into the other adiabatic region after $g=\hat{g}_2$ with $\zeta_s(\hat{g}_2)=\zeta_d$. Various excitation modes are left in this region because of the diabatic dynamics in the preceding impulse region. Among them, the topological defects survive for very long time after the quench. The KZM shows that the density of the topological defects $n$ obeys~\cite{revqkz1,revqkz2}
\begin{equation}
  n\propto R^{\frac{1}{r}}, \label{kzmtp}
\end{equation}
in which $r\equiv z+1/\nu$. Although the complete freezing of the evolution in the impulse region has been shown to be an oversimplified assumption~\cite{Keesling}, the KZM prediction Eq.~(\ref{kzmtp}) has been verified in various LGW phase transitions~\cite{Ulm,Pyka,Navon,Autti,Ko,Keesling}.
\begin{figure}
	\includegraphics[angle=0,scale=0.17]{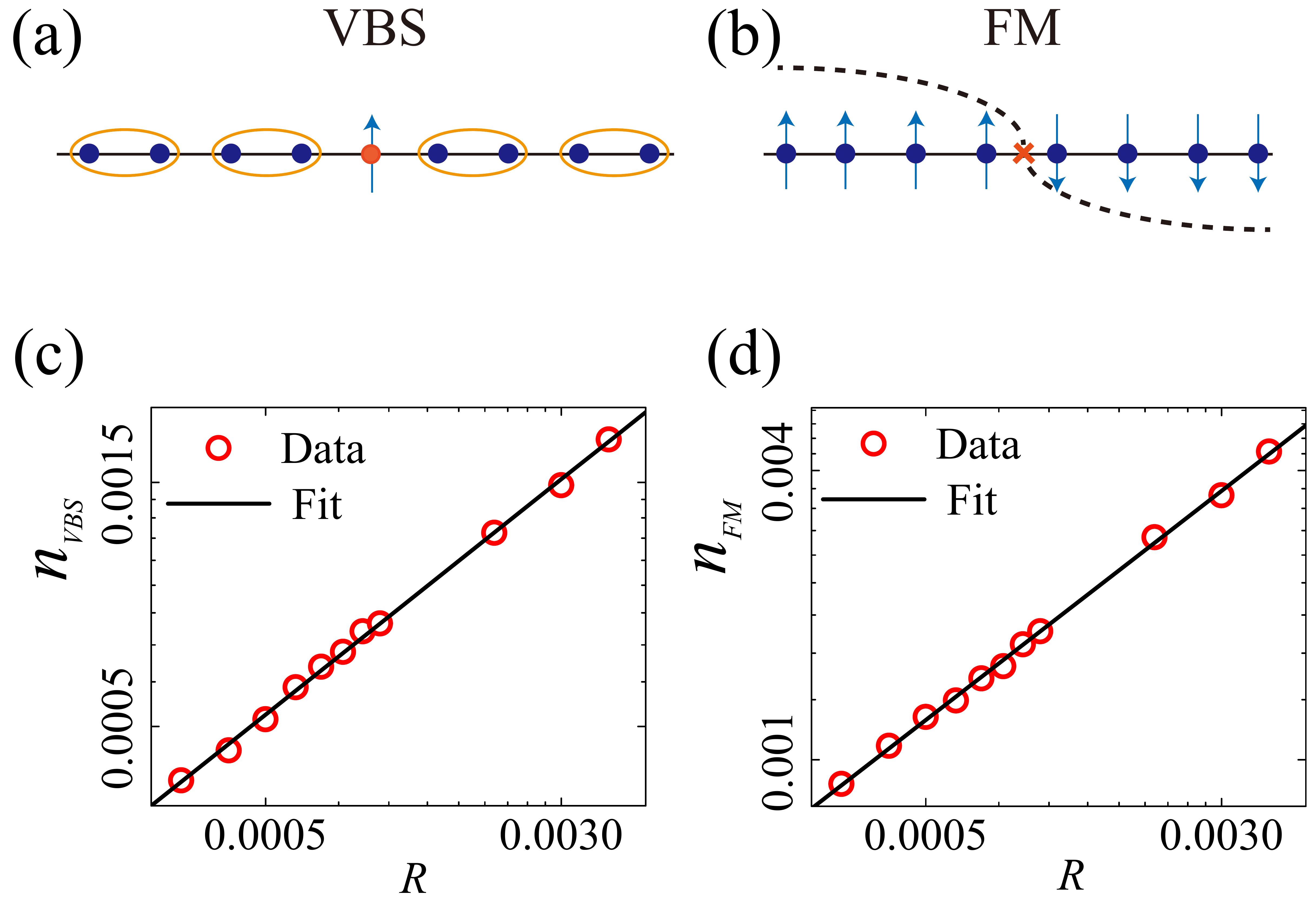}
	\caption{Topological defects in the VBS phase and the FM phase are shown schematically in (a) and (b), respectively. Under linear quench from the FM phase to the VBS phase, the density of the topological defects at the Majumdar-Ghosh point versus the driving rate is shown in (c). Power fitting shows that $n_{\rm VBS}\propto R^{0.6123}$. Similarly, Under linear quench from the VBS phase to the FM phase, the dependence of the density of the topological defects at $J_z=2$ on the driving rate is shown in (d). Power fitting shows that $n_{\rm FM}\propto R^{0.6124}$. Double-logarithmic coordinates are used in (c) and (d).}
	\label{figure2}
\end{figure}

Here, we examine the KZM in the $(1+1)$D DQCP of model~(\ref{modelham}). We at first consider the driving process from the FM phase to the VBS phase. According to the KZM, the topological defects emerge in the VBS phase. The ground states of the VBS order are of two-fold degeneracy, corresponding to two kinds of dimerized configurations. The topological defect therein is just the unpaired spin between two different ordered segments as shown in Fig.~\ref{figure2} (a). For the sake of simplicity, we measure the density of the topological defects at the Majumdar-Ghosh point with $J_z=1$~\cite{Majumdar}. In this way, the density of the topological defects is
\begin{equation}
n_{\rm VBS} =\left|\langle {\bf S}_i\cdot {\bf S}_{i+1}\rangle-\frac{1}{8}\right|, \label{tpvbs}
\end{equation}
in which $1/8$ is the average of the expectation value of ${\bf S}_i\cdot {\bf S}_{i+1}$ at each bond for the ground state.

We use the infinite time-evolving block decimation method to simulate the driven dynamics and calculate Eq.~(\ref{tpvbs}) for various driving rates~\cite{Vidal,SupM}. Figure~\ref{figure2} (c) shows the dependence of $n_{\rm VBS}$ on $R$. The power-law fitting gives $n_{\rm VBS}\propto R^{0.6123}$, confirming the KZM Eq.~(\ref{kzmtp}), in which $1/r\simeq 0.617$.

Similarly, we consider the reverse driving from the VBS phase to the FM phase. The topological defect in the FM phase is the domain wall between two segments of degenerate FM states, as shown in Fig.~\ref{figure2} (b). Quantitatively, the density of the topological defects reads~\cite{qkz1,qkz2}
\begin{equation}
n_{\rm FM} =  \left|\langle S^z_i S^z_{i+1}\rangle-\langle S^z_i S^z_{i+1} \rangle_G \right|, \label{tpfm}
\end{equation}
in which $\langle\rangle_G$ represents the expectation value in a ground state far from the critical point. By calculating Eq.~(\ref{tpfm}) for different $R$,  one finds that $n_{\rm FM}\propto R^{0.6124}$ as shown in Fig.~\ref{figure2} (d), confirming the KZM again. From these results, one concludes that the KZM is still applicable in the $(1+1)$D DQCP of model~(\ref{modelham}).

\begin{figure}
	\includegraphics[angle=0,scale=0.17]{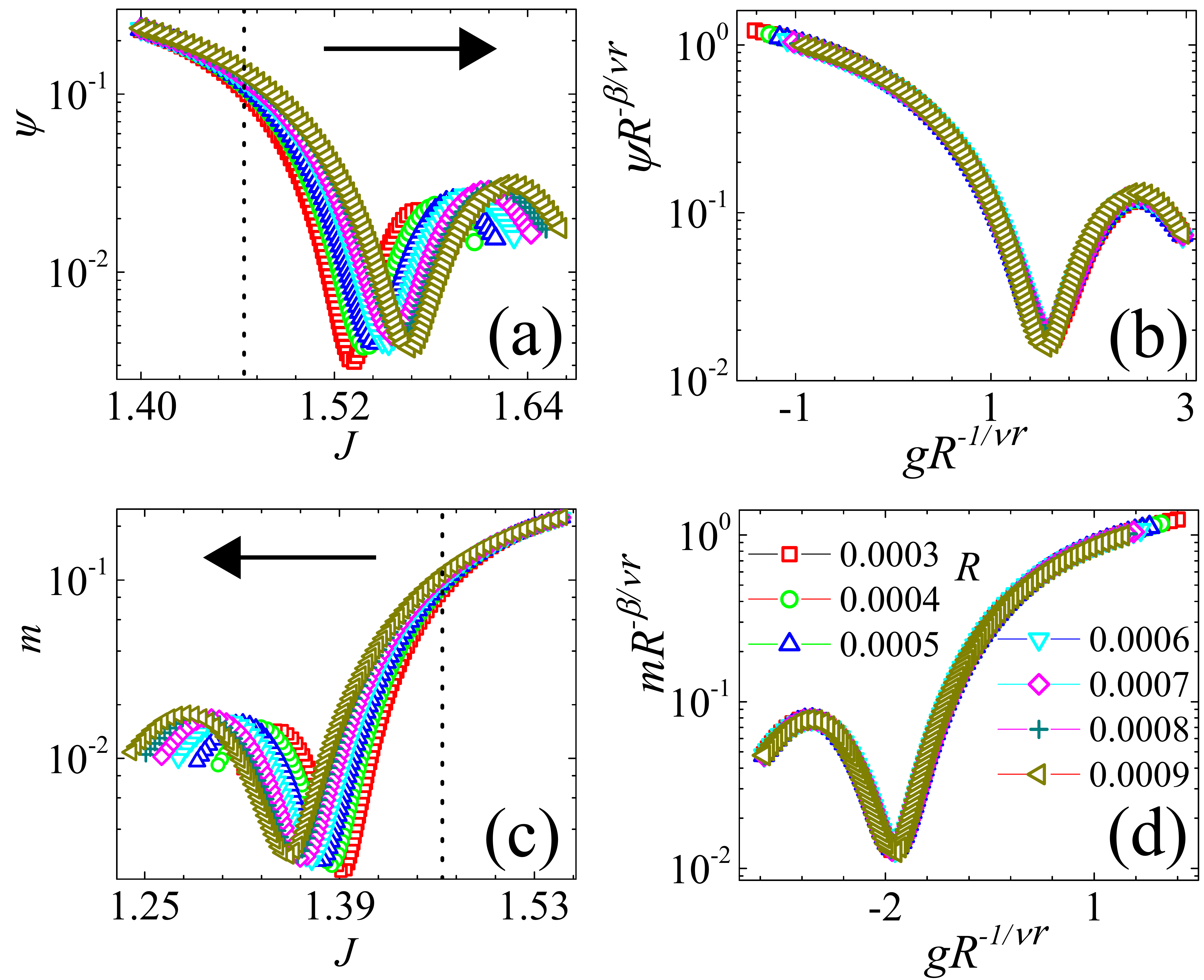}
	\caption{For increasing $g$, the evolution of the VBS order parameter with different driving rate indicated is shown in (a). The corresponding rescaled curves are shown in (b). Similarly, for decreasing $g$, the evolution of the FM order parameter with different driving tate is shown in (c). The corresponding curves after rescaling are shown in (d). Arrows in (a) and (c) indicate the driving direction. Vertical dotted lines in (a) and (c) mark the critical point. Logarithmic coordinates are used in vertical axes.}
	\label{figure3}
\end{figure}

{\it FTS in $(1+1)$D DQCP}.--- In the impulse region, the KZM states that the system does not evolve. However, it has been shown that this is an oversimplified assumption. The FTS improves the understanding of the driven critical dynamics by demonstrating that in this region the system evolves according to the characteristic time scale $\zeta_z\sim R^{-z/r}$~\cite{Zhong1,Zhong2}. In analogy to the finite-size scaling, the FTS theory shows that the evolution of the macroscopic quantities should satisfy the scaling forms with $\zeta_z$ in them. For example, the VBS order parameter $\psi$ for changing $J_z$ from the VBS to the FM phase should obey~\cite{Yin,Zhong1,Zhong2}
\begin{equation}
    \psi(g,R) = R^{\beta/\nu r} f_1 (g R^{-1/ \nu r}).
    \label{vbsfts}
\end{equation}
in which $f_i$ is the scaling function. Similarly, for the reverse driving the FM order parameter $m$ should satisfy~\cite{Yin,Zhong1,Zhong2}
\begin{equation}
    m(g,R) = R^{\beta/\nu r} f_2 (g R^{-1/ \nu r}),
    \label{mfts}
\end{equation}

\begin{figure}
	\includegraphics[angle=0,scale=0.17]{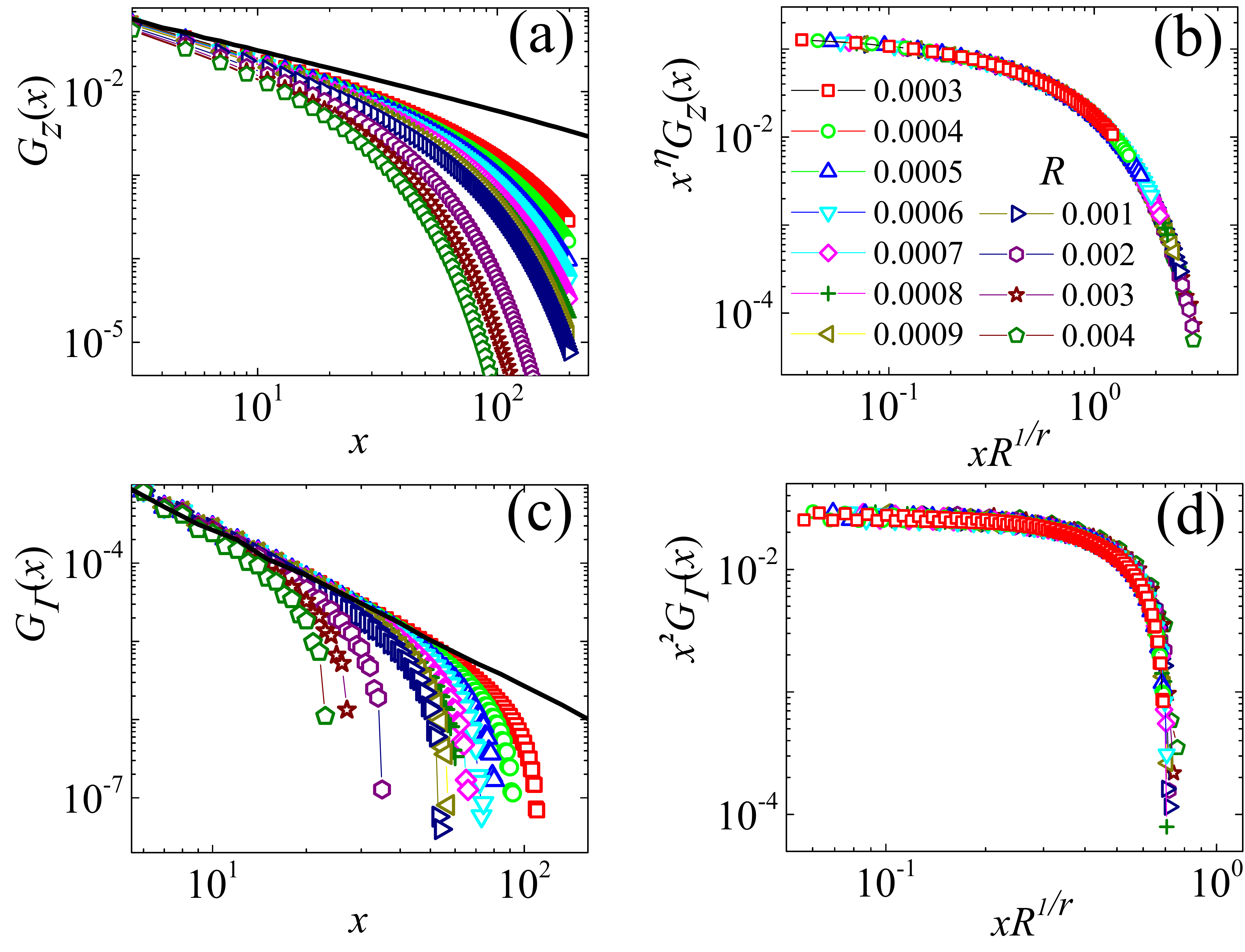}
	\caption{For various driving rates indicated, the spin-spin correlation $G_z$ before and after rescaling are shown in (a) and (b), respectively. Similarly, evolutions of the $xy$-dimer correlation $G_\Gamma$ before and after rescaling are shown in (c) and (d), respectively. Solid black lines in (a) and (c) show the power functions of $G_z$ and $G_\Gamma$, respectively, for $R=0$. Double-logarithmic coordinates are used.}
	\label{figure4}
\end{figure}

Besides these local operators, correlation functions were also measured in experiments~\cite{Keesling}. At the critical point, the scaling form of the FTS for $G_z(x)\equiv \langle S_i^z S_{i+x}^z\rangle$ is~\cite{Yin,Zhong1,Zhong2}
\begin{equation}
    G_z(x) = x^{-\eta} f_3 (x R^{1/r}),
    \label{corfts}
\end{equation}
with $\eta\simeq 0.68$~\cite{Jiang,HuangRZDQCP}. In addition, the emergent $O(2)\times O(2)$ symmetry gives an exact exponent in the $xy$-dimer correlation~\cite{HuangRZDQCP}, which under external driving should satisfy~\cite{Yin,Zhong1,Zhong2}
\begin{equation}
    G_\Gamma(x) = x^{-2} f_4 (x R^{1/r}),
    \label{cor2fts}
\end{equation}
according to the scaling analysis of the FTS.

Figure~\ref{figure3} (a) shows our numerical results of the evolution of the VBS order parameters for various $R$. After rescaling according to Eq.~(\ref{vbsfts}), one finds in Fig.~\ref{figure3} (b) that all curves collapse onto a single one. Similarly, as shown in Figs.~\ref{figure3} (c) and (d), for the reverse driving, all curves of the FM order parameters match with each other after rescaling according to Eq.~(\ref{mfts}). In addition, by noting that the critical exponents employed to rescale the FM and the VBS order parameters are the same, we verify the self-dual property of this DQCP~\cite{Jiang} from the nonequilibrium dynamics.

From Fig.~\ref{figure4}, one finds that both $G_z$ and the $G_\Gamma$ decay faster than power law. And the larger the driving rate, the faster the correlations decay. This demonstrates that the external driving imposes an effective length scale in the correlations. Their FTS forms are confirmed in Fig.~\ref{figure4} (b) and (d), respectively. These results verify the FTS theory in the $(1+1)$D DQCP of model~(\ref{modelham}). In addition, from Eq.~(\ref{cor2fts}), one finds that the dynamic scaling provides alternative evidences in detecting emergent symmetries in DQCP~\cite{HuangRZDQCP}.

{\it Discussion}.--- Here, we pursue the reason for the robustness of the KZM and the FTS. Although the field theories in the LGW phase transition and the $(1+1)$D DQCP are different, the scaling properties in low-energy levels are quite similar. For example, for both cases the expectation values in these low-energy levels are all controlled by the critical point at the ground state. In the driven dynamics, low-energy excitation modes are generated by the external driving and begin to occupy the low-energy excited levels owing to the diabatic process in the impulse region, and the proportion of these excitation modes depends on the driving rate. This gives rise to the KZM and the FTS. This picture should be applicable in both the LGW phase transition and the $(1+1)$D DQCP. Actually, the driven dynamics was also studied in topological phase transitions~\cite{Bermudez1,Bermudez2,Choi,Liou}, which are also beyond the LGW paradigm. These studies show that the KZM breaks down for the open boundary condition. The reason is that for these systems the structure of the energy levels is different from the usual LGW phase transition and the present $(1+1)$D DQCP owing to the appearance of the edge state~\cite{Bermudez1,Bermudez2,Choi,Liou}.

Phenomenologically, the $(1+1)$D DQCP is analogous to the $(2+1)$D DQCP in the sense that both of them happen between two ordered phases and exhibit emergent symmetries. Theoretically, various field theoretical descriptions of the $(2+1)$D DQCP can find their parallel counterparts in the $(1+1)$D case~\cite{Jiang}. These similarities indicate that our results could be generalized to the $(2+1)$D DQCP. However, the fractionization dynamics, such as the splitting of the spin wave into the spinons, is not included here. We note that a recent study has investigated the $(1+1)$D DQCP in a long-range interaction model~\cite{YangSB}. The deconfinement process can be studied therein. Moreover, in the $(2+1)$D DQCP, dangerously irrelevant variables may complicate the FTS analysis~\cite{Shao}.  Works on these projects are in progress.

{\it Summary}.--- Summarizing, in this paper we have taken a first study in the nonequilibrium critical dynamics of the DQCP. We presented a detailed study of the driven dynamics in a $(1+1)$D incarnation of the DQCP. This phase transition happens between the FM phase and the VBS phase and is beyond the LGW paradigm. By studying the scaling of topological defects in both the FM phase and the VBS phase after the quench, we have confirmed the KZM. Moreover, we have also shown that the evolutions of macroscopic physical quantities, including the FM and the VBS order parameters and the correlation functions, satisfy the FTS theory. In particular, we have also explored the nonequilibrium dynamic version of the correlation scaling constrained by the emergent symmetry. In this way, we have generalized the KZM and the FTS into the $(1+1)$D DQCP. Our conclusions can be generalized to the $1$D fermion systems, in which the $(1+1)$D DQCP between a charge-density-wave phase and a dimerized phase was also found~\cite{Sandvik04}. Moreover, although exploring the real-time dynamics of the $(2+1)$D DQCP still needs a tremendous amount of work, it is expected that some of our results should be applicable therein with some modifications. Recently, reliable platforms based on the Rydberg-atom systems were realized to manipulate and detect the critical dynamics in spin systems with programmable interactions~\cite{Keesling}. Our results could be examined experimentally in these systems.

{\it Acknowledgments}.--- R. Z. H is supported by the Strategic Priority Research Program of Chinese Academy of Sciences (Grant XDB28000000). S. Y. acknowledge the support by the startup grant in Sun Yat-Sen University.

\begin{widetext}

\section{Supplemental Material}
\renewcommand{\theequation}{S\arabic{equation}}
\setcounter{equation}{0}
\renewcommand{\thefigure}{S\arabic{figure}}
\setcounter{figure}{0}
\renewcommand{\thetable}{S\arabic{table}}
\setcounter{table}{0}
\newcommand{\eqs}[1]{\begin{equation}\begin{split}#1\end{split}\end{equation}}

\section{S-1. Numerical method and calculation set up}
In this work we have utilized infinite matrix product state(MPS)~\cite{Cirac,Schollwock} based numerical methods to study the driven dynamics. As shown in Fig.~\ref{Fig_MPS}(b), a MPS is a generic class of quantum many body state which is represented as multiplication of matrices, which is the underlying ground state structure of the famous density matrix renormalization group method~\cite{ref_dmrg}. At each site there is a set of matrices, the dimension $\chi$ of which controls the upper bound of entanglement in a MPS. It has been shown MPS based methods can accurately describe ground states of gapped quantum lattice models and generally generic low entangled states in 1D. In recent years MPS based methods have been widely used to study both the static and dynamic properties of (quasi)$1$D quantum systems~\cite{Cirac2,Schollwock2}.

\begin{figure}[bp]
\includegraphics[angle=0,scale=0.3]{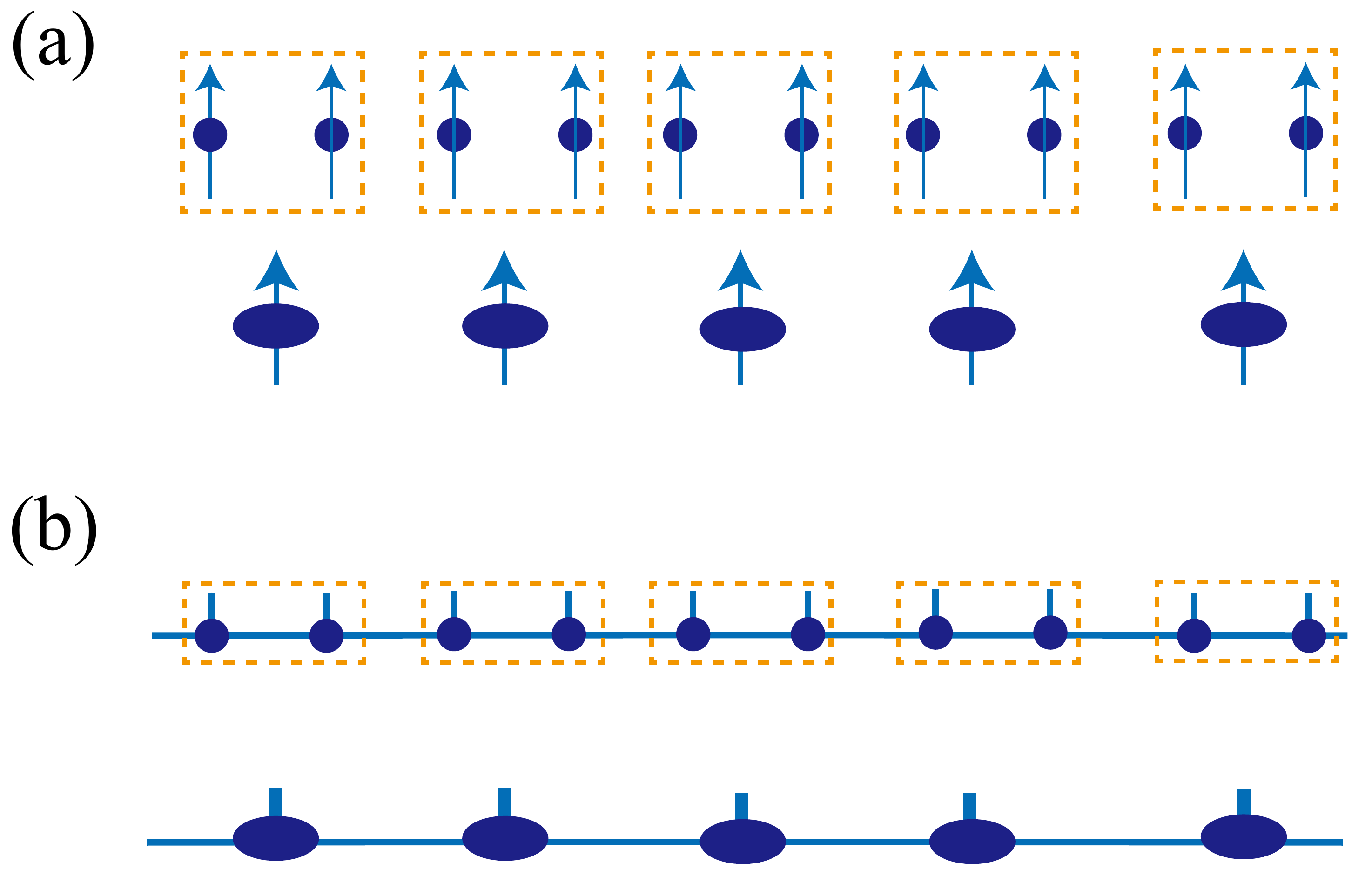}
  \caption{Schematic figure of a quantum chain (a) and MPS (b) after combining each two neighbor spins together.}
\label{Fig_MPS}
\end{figure}

To study the driven dynamics of a quantum lattice model, one first needs to calculate the ground state $\psi(0)$ at $g(0)$, then carry real time evolution calculation to obtain $\psi(t)$. Since $g(0)$ is far away from the quantum critical point, the system has a large energy gap and $\psi(0)$ is a low entangled state. Moreover it has already been shown the entanglement entropy is very limited in the driven dynamics due to the external driving field~\cite{Zhongee}. So that both $\psi(0)$ and $\psi(t)$ have limited entanglement and can be accurately described using the MPS.

\subsection{Ground state calculation}
We use an infinite MPS to variationally optimize the ground state. To be consistent with both the FM and the VBS phase, a two site unit-cell in the MPS was used during the calculation. In order to obtain an accurate MPS wave function for the ground sate $\psi(0)$, we adopt the recent proposed tangent space MPS variational method\cite{ref_vmps}. One can directly calculate the energy gradient in the given $\chi$ MPS sub-manifold, and use that to optimize the ground state energy until it becomes convergent.

\subsection{Real time evolution}
After obtaining the ground state $\psi(0)$, we use the infinite time-evolving block decimation(TEBD) method~\cite{ref_i_tebd} to carry the real time evolution. Upon the evolution for a very small time slice $t$, one first needs to decompose the full time evolution operator $\textrm{exp}(-i\mathcal{H}t)$ into local evolution ones via Suzuki-Trotter decomposition and then update the matrices in the MPS after imposing the evolution operator.

However an extra difficulty lies in the present model we study. The next nearest neighbor(NN) interactions in the Hamiltonian can complicate the higher order Suzuki-Trotter decomposition. To avoid the complex decomposition and its following evolution procedure, we combine every two neighbor sites into a single one, as shown in Fig.~\ref{Fig_MPS}. The dimension for the local Hilbert space raises from 2 to 4. In this way, the NN term becomes nearest neighbor again and the conventional TEBD procedure can be used to carry the real time evolution.

\subsection{Convergence and Stability}
We have taken the following setup for the MPS simulation to ensure the convergence and stability. During the ground state calculation, the energy gradient with respect to the matrix elements was required to be smaller than $10^{-12}$ to obtain converged wave function. Moreover we have taken several random initial MPS for the variational calculation to find the best approximated state, so that local minimum can be avoided. During the time evolution calculation, a small single time slice $t=0.01$ and the fourth order Suzuki-Trott decomposition were used. This ensures the Trotter error in one single time step is about $10^{-10}$. The largest $\chi$ for the MPS in our calculation is $200$, and larger $\chi$ has been used to check the convergence of our results. Moreover, a pseudo-inverse of the singular value decomposition spectra was used, in which extremely small elements have been dropped, to ensure the stability.

\begin{figure}[bp]
\includegraphics[angle=0,scale=0.2]{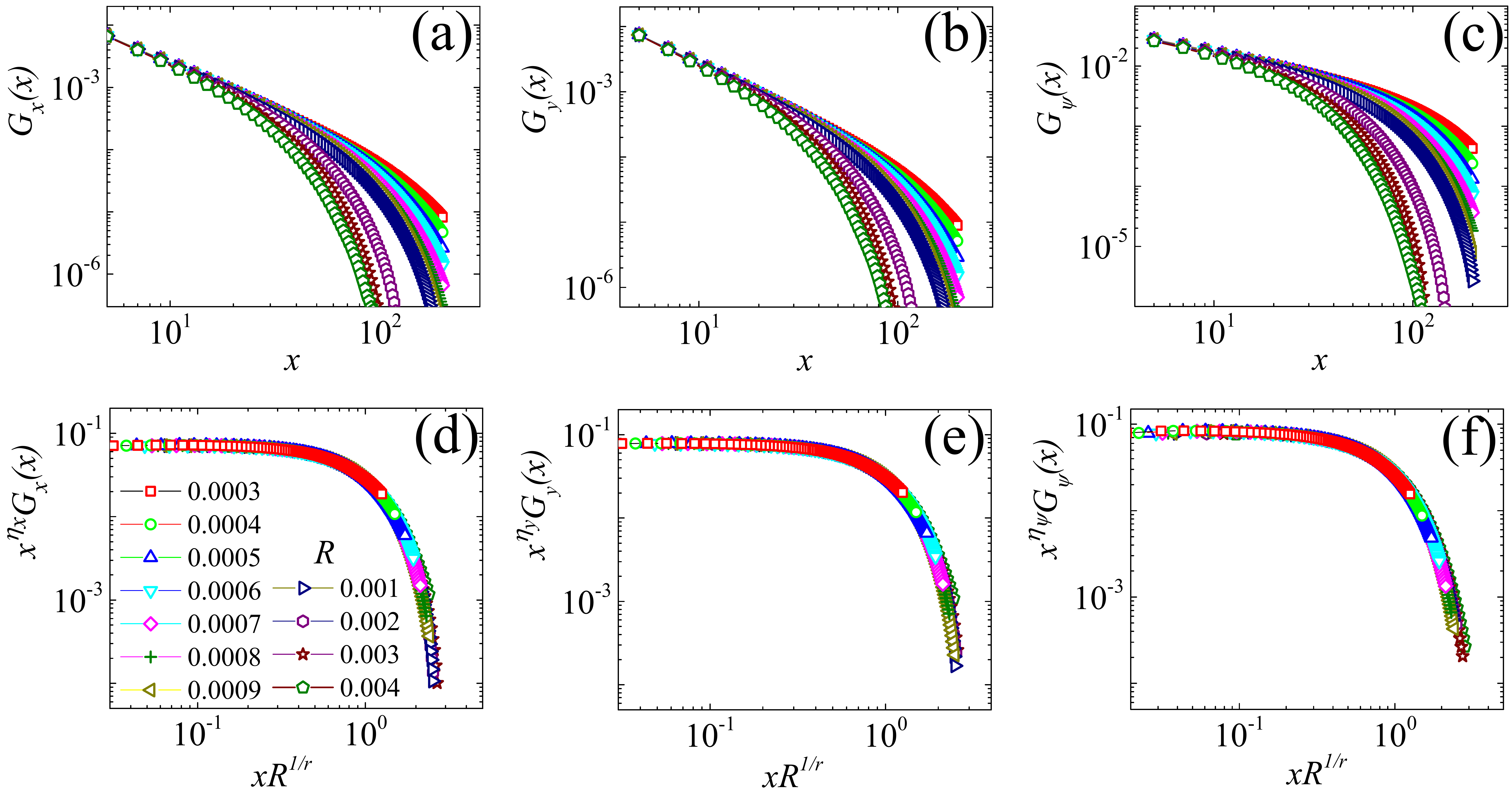}
  \caption{(a-c) The correlation functions defined in Eq.~\ref{seq:correlations} for various driving rates before (a-c) and after (d-f) rescaling. The exponents $\eta_x = \eta_y = 1/\eta$ and $\eta_\psi = \eta$.}
\label{sfig_correlation}
\end{figure}

\section{S-2. Correlation function}
Besides the spin-spin and the xy dimer correlation functions discussed in the main text, we also examine other correlation of scaling operators
\eqs{
G_x(r)&=\langle S_i^xS_{i+r}^x\rangle,\\
G_y(r)&=(-1)^r \langle S_i^yS_{i+r}^y\rangle,\\
G_\Psi(r)&=(-1)^r\langle \Psi_{i}\Psi_{i+r}\rangle,
\label{seq:correlations}}
where $\Psi_i= {\bf S}_i\cdot{\bf S}_{i+1}$ is the dimmer operator. According to the finite time scaling (FTS) theory, their leading contribution satisfy similar scaling behavior
\eqs{
G_x(r) & = x^{-\eta_x} f_4 (x R^{1/r}),\\
G_y(r) & = x^{-\eta_y} f_5 (x R^{1/r}),\\
G_\Psi(r) & = x^{-\eta_\Psi} f_6 (x R^{1/r}),
\label{seq:scaling}}
where $\eta_x = \eta_y = 1/\eta$ and $\eta_\Psi = \eta$.

The numerical results from the MPS calculation for the various correlations defined in Eq.~\ref{seq:correlations} are shown in Fig.~\ref{sfig_correlation} (a-c). Their scaling behavior is confirmed in Fig.~\ref{sfig_correlation} (d-f). These results verify the FTS theory again. Moreover, from the result that the dimer-dimer correlation $G_\Psi$ and the spin-spin correlation $G_z$ has the same exponent, we confirm the self-duality of the $1+1$D DQCP again.

\end{widetext}

\end{document}